## Electrically driven dipole layer formation

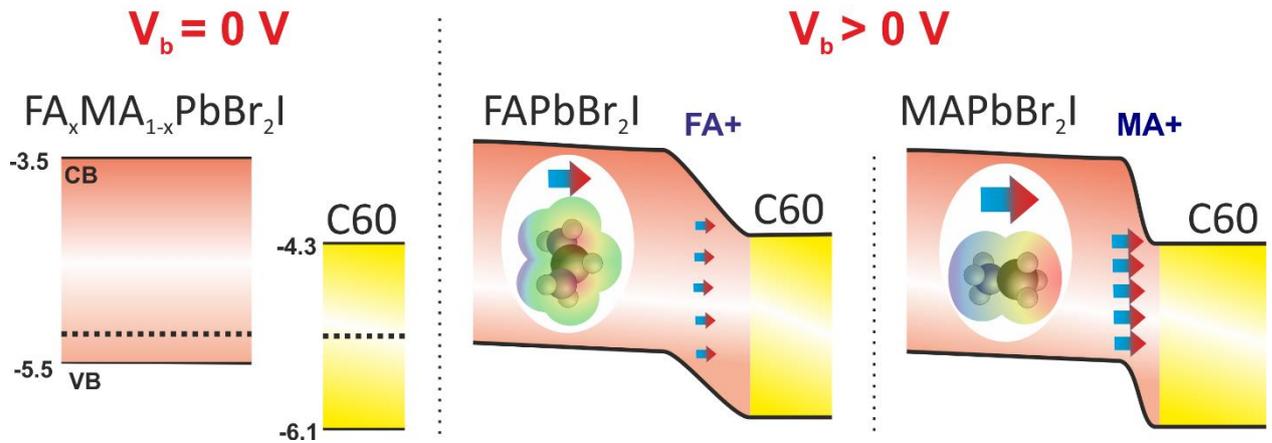

Dipolar cation accumulation at interfaces of perovskite light emitting solar cells

D.S. Gets[1], G.A. Verkhogliadov[1], E.Y. Danilovskiy[1], A. I. Baranov[2], S.V. Makarov[1], A.A. Zakhidov[1,3].

[1] ITMO University, Department of Nanophotonics and Metamaterials, Lomonosov Str. 9, Saint-Petersburg 191002, Russia
[2] St Petersburg Academic University, ul. Khlopina 8/3a, St Petersburg 194021, Russia
[3] Alan G. MacDiarmid NanoTech Institute, Department of Physics, University of Texas at Dallas, Richardson, TX 75083, USA
E-mail addresses: dmitry.gets@metalab.ifmo.ru (D.S. Gets), zakhidov@utdallas.edu (A.A. Zakhidov).

## Abstract

Ionic migration in organo-halide perovskites plays an important role in operation of perovskite based solar cells and light emitting diodes. Despite the ionic migration being a reversible process, it often leads to worsening of perovskite based device performance, hysteresis in current-voltage characteristics, and phase segregation in mixed halide perovskites being as the most harmful effect. The reason is in dynamical band structure changes, which controllable engineering would solve one of the biggest challenges for development of light-emitting solar cells. Here we demonstrate controllable band bending due to migration of both cation and anion ions in mixed halide perovskite devices. The band structure rearrangement is demonstrated in light emitting solar cells based on the perovskite with organic cations methylammonium ($MA^+$) and formamidinium ($FA^+$), possessing non-zero dipole momentum of 2.29 and 0.21 Debye, respectively, and with PEDOT:PSS and C60 transport layers having a high barrier of 0.8 eV for charge injection. Under applied external voltage $MA^+$ and $FA^+$ cations move towards the electron transport layer and form a dipole layer at the perovskite/electron transport interface, which lowers threshold voltage for electroluminescence down to 1.7 V for $MAPbBr_2I$ and 2.6 V for $FAPbBr_2I$, whereas monohalide perovskite $MAPbBr_3$ does not demonstrate such behavior. This ability to *in-situ* change the device band structure paves the way developing of dual-functional devices based on simple design. It also makes mixed halide perovskites more flexible than mono halides ones for developing different optoelectronic devices without the use of special types of work function modifying transport materials.

## Introduction

Organo-halide perovskites (OHP) is an encouraging family of organic-inorganic materials for developing highly efficient solar cells (SCs) and light emitting diodes (LEDs) [1, 2]. OHP have a high absorption coefficient, high exciton binding energy, wide band gap tunability, and solution processing and many more [3-7]. SCs based on perovskite reached efficiencies of SCs produced by well-established technologies like silicon [8], and perovskite LEDs demonstrate high efficiencies, narrow luminescence line width and high color rendering indexes [9]. But OHPs demonstrate some unwanted effects like ionic migration which tremendously affects the device performance [10, 11]. Ionic migration is induced either by illumination [12] or by passing the high current [13] through the device and most clearly manifested in mixed halide perovskites where it results in anion segregation [10, 12]. Perovskite ions and their vacancies have low activation energies, including mobility even at room temperature and for the long time, mnaking the ionic migration one of the biggest problems for perovskite devices [14-19]. The ionic migration strongly affects the device performance and results in hysteresis of PV parameters [20]. It clearly seen in current-voltage characteristic (J-V) of perovskite based solar cells where efficiency depends on the direction of voltage sweep. This unwanted behavior can be slow down by different ways like using multiple cation composition, producing perovskite nanocrystals, increasing perovskite grain size and interface stabilization [21-25].

Despite the negative effects of ionic migration, it can also demonstrate some positive effects. For example, under light illumination or application of voltage perovskite ions move towards electron and hole transport layers (ETL and HTL) and thus forming the p-i-n structure inside the perovskite layer [18, 26-29]. This p-i-n structure is responsible for temporal performance enhance of solar cells, which is known as light induced self-poling effect [26], and also it makes possible creation of dual-functional devices – light emitting solar cells (LESC) [30-32].

Latest experimental investigations of ionic migration in MAPbI$_3$ shows that the anion ion has greatest diffusion coefficient and cation ion stays almost intact [33, 34], which means that p-i-n structure inside the perovskite layer is formed mostly due to halogen ion migration. Possibility of p-i-n structure formation due to ion migration inside the perovskite layer aids to the development of the LESC.

The main problem of LESC is the absence of a mechanism to *in-situ* change the device band structure for certain working regime, LED or SC. Therefore, dual functional device will have additional losses due to the poor band alignment between the transport and emission layers when working in a reciprocal regime. Hence, creating LESC with high efficiencies in both working regimes is a challenge. Usually to overcome this potential barrier there is a need to insert additional layers to adjust work function of electrode [31, 32]. There are examples of differnet LESCs [35-37] the perovskite LESC based on standard design with PEDOT:PSS as HTL and C$_{60}$ as ETL can work as a LESC despite high potential barrier (≈0.8eV) between the perovskite layer and ETL [30]. And the device within simple design can demonstrate low threshold voltage ($V_{th}$) and relatively good device performance with electroluminescence efficiency up to 0.04% (EQE$_{EL}$) and photoconversion efficiency (PCE) up to 3-4% [30]. While in MAPbBr$_3$-based LESC, which have high potential barrier between electrode and perovskite polyelyctrolyte ETL was used and thus LESC has the PCE ≈ 1% and EQE$_{EL}$ ≈ 0.12% [31], efficiencies. And in the case of MAPbI$_3$-based LESC low work function electrode was used and device have PCE ≈ 12% and EQE$_{EL}$ ≈ 0.04% [32].

In this work, we demonstrate controllable real time band bending of the perovskite LESC based on mixed halide perovskites due to dipole layer formation from perovskite cation ions. Mixed halide perovskites demonstrate more pronounced ionic migration than monohalide ones due to the movement of both cation and anion ions. The difference in ionic migration is demonstrated by a high change in the LED threshold voltage values after prebiasing the device at different biasing voltages ($V_b$). The prebiasing results in migration of Br$^-$ and I$^-$ ions towards the PEDOT:PSS/perovskite interface and MA$^+$ and FA$^+$ molecules migrate towards the perovskite/C$_{60}$ interface form the accumulation layers. The presence of these accumulation layers at the perovskite interfaces leads to p-i-n structure formation inside the perovskite layer, and, since, molecules of perovskite organic cations (MA$^+$ and FA$^+$) have non-zero dipole momentums (2.29 and 0.21 Debye) [38], the cation accumulation layer at the interface will bend device band structure more efficiently. This leads to improvement of charge injection with consequent $V_{th}$ lowering down to 1.7V.

*Device Fabrication*

For the device fabrication the following scheme of functional layers ITO/PEDOT:PSS/FA$_x$MA$_{1-x}$PbBr$_2$I/C$_{60}$/LiF/Ag was chosen. The device functional layers were subsequently deposited onto ITO covered glass substrates and glass substrates with ITO were cleaned in an ultrasound bath in deionized water, acetone and isopropyl alcohol consequently. Dried substrates were exposed to UV irradiation (189, 254nm) for 900s. Water dispersion of PEDOT:PSS 4083 was used as a HTL, it was filtered through PTFE 0.45 syringe filter and deposited by spin coating. After spincoating the film was annealed on a heating plate for 10min at 150°C. A photoactive layer based on FA$_x$MA$_{1-x}$PbBr$_2$I was prepared by the consequent dissolving methylammonium iodide (MAI, DyeSol) and lead (II) bromide (PbBr$_2$, Alfa Aesar "Puratronic" 99,999%) in a mixture of dimethylformamide (DMF) and dimethyl sulfoxide (DMSO) DMF:DMSO (7:3) respectively. The solutions were stirred overnight at room temperature. The acquired perovskite inks were deposited by the single step solvent engineering technique on top of HTL in two step spin-cycle inside the glovebox system with nitrogen atmosphere. Diethyl ether was used as an antisolvent and was slowly dripped on the rotating substrate in 10 s past the increase from 1000 to 3000 rpm. The acquired perovskite films were subjected to vacuum annealing for 1 minute with consequent annealing on the heating plate at 100°C for 10 minutes. $C_{60}$ layer was used as ETL. $C_{60}$ as well as LiF and Ag layers were deposited by thermal evaporation.

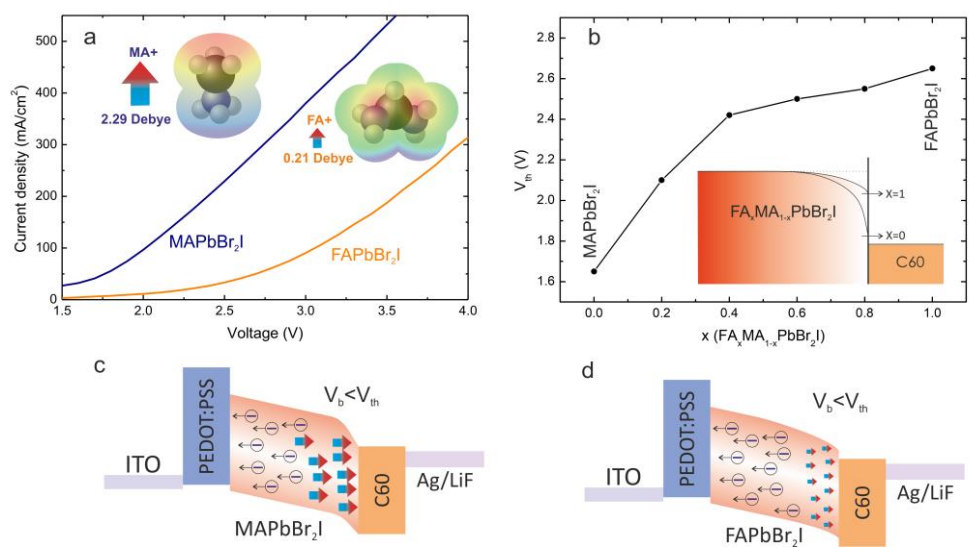

**Figure 1. Ion migration.** **a** J-V curves of MA- and FA-based perovskite devices after previasing at 1.5 Volts. Molecules of MA (d = 2.29 Debye) have greater dipole momentum than FA (d = 0.21 Debye) and therefore MA-based devices have stronger band bending and, consequently, lower $V_{th}$ than FA-based devices. **b** dependence of $V_{th}$ on cation composition of mixed halide perovskite. Values of $V_{th}$ were obtained from linear approximation of J-V characteristics after several biasings of the devices at 1.5V for 1 minute in dark. The inset demonstrates probable band bending for devices based on MAPbBr$_2$I (x = 0) and FAPbBr$_2$I (x = 1). **c** and **d** Cation ion moves towards $C_{60}$ layer and iodine and bromide ions moves towards PEDOT:PSS layer. Rearranged device band structure due to dipole layer formation of LESC based on MAPbBr$_2$I and FAPbBr$_2$I.

### Results and discussion

To investigate beneficial effect of ionic migration and p-i-n structure development inside the perovskite layer for band bending two perovskite organic cations FA$^+$ and MA$^+$ was chosen. The MAPbBr$_2$I perovskite has a little bigger band gap than FAPbBr$_2$I and FA cation offers greater stability against segregation than MA cation. For precise investigation of dual-functional devices performance of upon cation composition in mixed halide perovskite devices with cation FA$_x$MA$_{1-x}$PbBr$_2$I were created (x = 0, 0.2, 0.4, 0.6, 0.8, 1). Obtained perovskite devices were tested for solar cell performance and they all demonstrated relatively good PV characteristics ($V_{OC} \approx 0.85$V, $J_{SC} \approx 9$mA/cm$^2$, FF ≈ 50%, PCE ≈ 3%) regardless cation composition (Fig. S1). Study of the LED device regime performance and p-i-n structure formation was conducted by a slightly modified procedure suggested earlier [30].

LESC is a device capable of working in two reciprocal regimes as a light emitting diode (LED) and a solar cell (SC). Generally, LEDs and SCs perform reciprocal functions of converting electrical power to the light and vice-versa. Although these devices share similar designs, their designs are tuned in a specific way to maximize performance of the primary function, and therefore, the reciprocal function is greatly suppressed by design due to additional losses because of high band mismatch. Therefore to create an efficient LESC a way to adjust band diagram for specific working regime should be found. However, dual-functionality in OHP devices was realized by using of low work function electrode material [32], a special transport material with polyelectrolyte properties [31], or ion movement [30]. The first two approaches are based on modulation of transport material properties. In the first case, the potential barrier for charge injection was overcome due to initially tuned device structure by using of low work function electrode material barium [32]. In the second case, the potential barrier was lowered by using polyethylenimine or pre-doped polyethylenimine (PEI or PEIBim$_4$) as an ETL [31]. It is important to note that PEI demonstrates polyelectrolyte properties, which offers internal dipole formation, and it is usually used as a universal buffer layer to lower electrode work function [39]. The presence of dipoles at the interface MAPbBr$_3$/ETL bends the band structure and, in turn, the potential barrier for charge injection also lowers, allowing for relatively good PV and LED performance in the same device. The third approach utilizes ion movement inside the perovskite layer [30]. Design used in these devices suits SC operation, but for LED operation it is highly unoptimized. The main problem is the high potential barrier for electron injection (≈0.8eV) between perovskite and ETL that makes this device design highly undesirable for LED realization. In this case under applied external voltage perovskite ions move towards perovskite-transport layer interfaces and form p-i-n structure inside the perovskite layer. This p-i-n structure aids in dual functionality. Recent investigations of halide diffusion in MAPbI$_3$ perovskite demonstrated high mobility of the I$^-$ ion (D ≈ 10$^{-9}$ cm$^2$s$^{-1}$) and extremely low mobility of the MA$^+$ ion (D ≈ 10$^{-12}$ cm$^2$s$^{-1}$) [19, 20, 27, 28]. Therefore, the p-i-n structure in perovskites forms only due to I$^-$ and Br$^-$ ions and its vacancies and it seems, which it is not enough to achieve high band structure bending in order to overcome potential barriers and enhance charge injection efficiency in the reciprocal working regime without using of any special electrode work function modifying layers.

Each LESC device was subjected to a sequence of biasing cycles at a certain biasing voltage ($V_b$ = 1, 1.5, 2 V etc.) for 1 minute in the dark and was immediately followed by J-V characteristic measurement from $V_b$ up to 3–4V (Fig. 1a). Applying the external electric field to the device leads to two processes: first, movement and accumulation of perovskite ions at the interfaces perovskite/ETL and perovskite/HTL, and second, current induced halide segregation [13]. An excess of perovskite ions at interfaces leads to p-i-n structure formation inside the perovskite layer since additional ions at interfaces serve as dopants [26, 18, 41]. Higher $V_b$ should result in higher ion density at the interface, which in turn results in strongly pronounced p-i-n structure. However, ionic migration as well as p-i-n structure formation and segregation are temporal effects, and removing the external electric field leads to reverse redistribution of ions inside the perovskite layer and p-i-n junction disappearance. Therefore, to achieve low $V_{th}$ the device must always be under applied external voltage. Along with J-V, we also measured the electroluminescence (EL) spectra of the LED (Fig. S2), which amplitude behavior supports J-V characteristics. $V_{th}$ values were deduced from zero crossing of J-V curve linear approximation (Fig. S2).

Fig. 1a demonstrates J-V characteristics of MA- and FA-based perovskite devices after several pre-biasings at 1.5V. The J-V characteristics demonstrate high difference in position of "knee". Obtained $V_{th}$ values for MA-based ($V_{th}$ ≈ 1.68V) and FA based ($V_{th}$ ≈ 2.62V) perovskites demonstrate high difference ($\Delta V_{th}$ ≈ 1V), which reflects the difference in charge injection into perovskite layer for these cations. For clarification the origin of such high difference between $V_{th}$ of MA- and FA-based perovskite devices we conducted the same sequence for different FA$_x$MA$_{1-x}$ (0≤x≤1) cation composition of perovskite. Fig. 1b demonstrates V$_{th}$ dependence for different cation compositions FA$_x$MA$_{1-x}$ (where x = 0, 0.2, 0.4, 0.6, 0.8, 1) after pre-biasings at V$_b$ = 1.5V. V$_{th}$ values obtained J-V characteristics demonstrate non-linear behavior upon cation FA$_x$MA$_{1-x}$ (0≤x≤1) composition (Fig. 1b). Also we investigated influence of V$_b$ value on V$_{th}$ value, Fig. 2a shows J-V curves of FA-based LESC measured after pre-biasing at different V$_b$, the black line corresponds to the J-V characteristic after pre-biasing at 1V, the blue line to pre-biasing at 1.5V and the orange line after the pre-biasing at 2V. The increase of V$_b$ leads to moving of the J-V characteristic "knee" to the low voltage region along with lowering of V$_{th}$ value (Fig. 2a). Each V$_b$ has its saturated value of V$_{th}$ and than the longer time the device is subjected to the pre-biasing by V$_b$ application results in more pronounced and shifted to the low voltage region "knee" of the J-V characteristic (Fig. S3) as well as attains saturated V$_{th}$ value (Fig. S4).

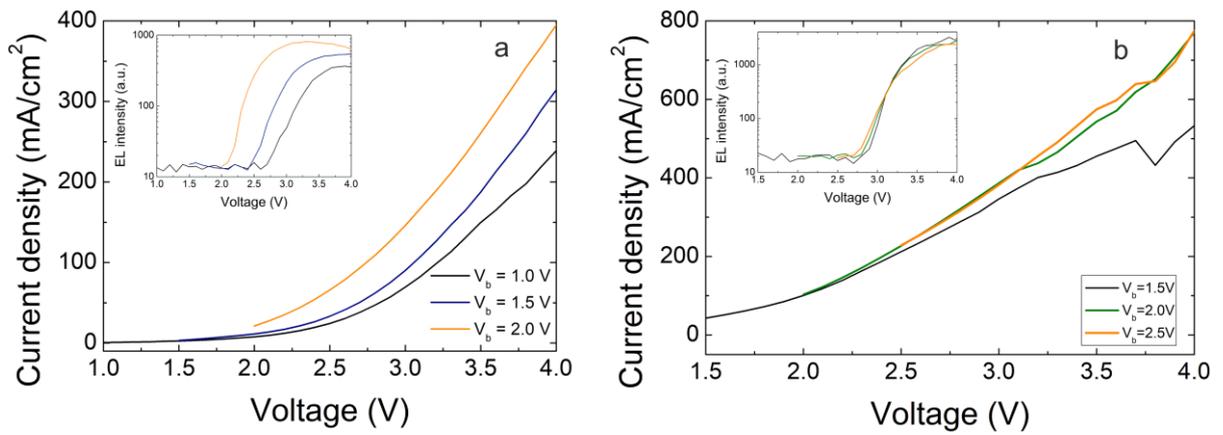

Figure 2. Biasing the mixed halide and monohalide perovskite devices. a J-V characteristics of the FAPbBr$_2$I-based device after different pre-biasing voltages (V$_b$ = 1, 1.5, 2V). Inset shows the behavior of EL amplitude after biasing under different V$_b$. b J-V characteristics of MAPbBr$_3$ based device after pre-biasing at different voltages. Inset demonstrates corresponding EL peak intensity after pre-biasing at different voltages.

Moreover, increase of V$_b$ also leads to higher amplitude of EL and as well as EL ignition occurs at the low voltage region, which reflects better charge injection conditions (inset in Fig. 3a). Devices with different cation compositions demonstrate different V$_{th}$ values after several pre-biasing cycles at equal V$_b$. This behavior of V$_{th}$ dependence and high total difference is quite surprising. We also recorded the optical power of devices in LED regime and external quantum efficiency, where the FA-based perovskite is much lower than MA-based (Fig. S5). While having similar characteristics to PV performance, LED emission line did not demonstrate any odd behavior, and all devices demonstrated an EL in the red region with dependence of peak position on FA$_x$MA$_{1-x}$ (0≤x≤1) composition (Fig. S6). Anion segregation induced by high current injection produces domains enriched with iodine ions inside the perovskite layer, and these domains control the EL.

It is important to note that, there is a high band mismatch between the perovskite layer and C$_{60}$ (Fig 1.), but the device is still capable of demonstrating a relatively high LED efficiency [30] and low V$_{th}$. The difference in LED efficiency and in V$_{th}$ values points to different conditions for charge injection into perovskite layer. MA-based perovskite offers better charge injection conditions than the FA-based. It turns out that the p-i-n structure in MA- and FA-based peroskites is different but in accordance with DFT calculations of defects in perovskite MAPbI$_3$ [41], interstitial iodine (I$_I$) ion, and MA vacancy (V$_{MA}$) produce an acceptor level near the valence band. At the same time, iodine vacancy (V$_I$) and the interstitial MA molecule (MA$_I$) produce a donor level near the conduction band [41], and almost the same situation occurs with FAPbI$_3$ [42], moreover, position of these type of defects does not change much in different perovskites [43, 44]. Therefore, the p-i-n structure in both cases should be similar, yet but experimental results shows the opposite. This contradiction points to additional effects influencing p-i-n development. Thus, we looked at properties of cation molecules MA and FA. Molecules of these cations have different and high dipole momentums of 2.29 and 0.21 Debye, respectively [38], and since MA molecules have a greater dipole momentum than FA molecules, MA-based device band bending will be stronger and their V$_{th}$ lower than FA-based ones (Fig. 1b).

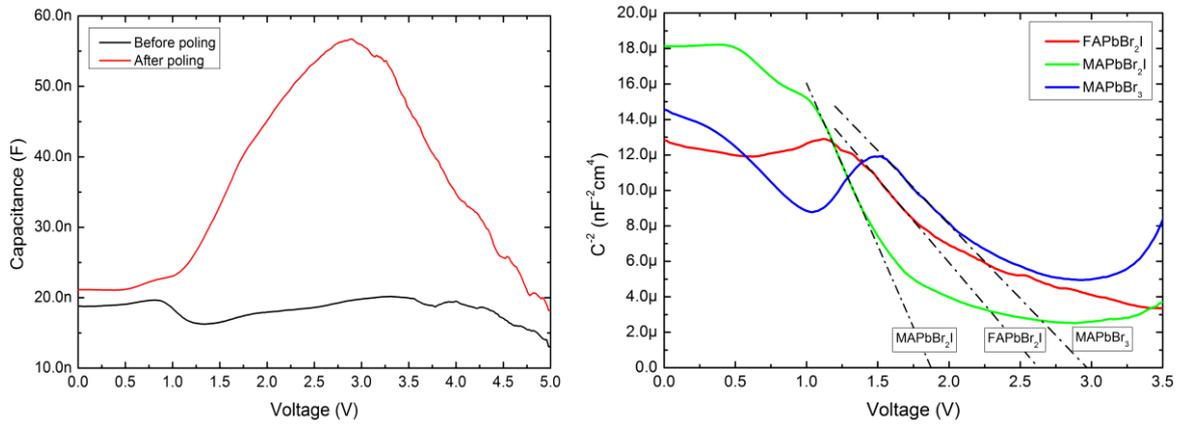

**Fig. 3. C-V characteristics of perovskite dual functional devices**. **a** C-V characteristics before (black line) and after (red line) prebiasing the device. **b** Mott–Schottky curves of prebiased dual functional devices based on MAPbBr$_2$I (green line), FAPbBr$_2$I (red line), MAPbBr$_3$ (blue line). Linear approximation gives V$_{bi}$ of devices.

To verify our hypothesis we conducted the same cycle of biasing on monohalide perovskite devices by investigating MAPbBr$_3$-based devices in same device design (Fig. S7). Fig. 2b demonstrates obtained J-V, EL-V characteristics and EL spectra. J-V characteristics of MAPbBr$_3$-based device, which do not change much event after biasing at high pre-biasing voltages (V$_b$ = 2.5 V). The absence of the shift in J-Vs (Fig. 3b) and EL (inset in Fig 3b) after biasing at V$_b$ = 1.5, 2.0, 2.5V corresponds to absence of MA$^+$-ion movement under applied voltage, otherwise these characteristics would demonstrate considerable shift to lower voltages and increase of EL intensity in a sequence of measurements (inset in Fig. 2b). Although there is a small shift in J-V and EL-V characteristic (Fig. 2b and inset in Fig. 2b) but in can be attributed to movement of Br$^-$ ions and formation of weak p-i-n structure due to movement of Br-ions and it's vacancies. Such behavior of J-V characteristics, V$_{th}$ value and electroluminescence in agreement with investigations of halide diffusion in MAPbI$_3$, in which molecule of MA almost doesn't move, but atoms of I have considerable diffusion coefficient [33]. Meaning that, the ionic migration of only halide ion is not suffisient to effectively bend perovskite device band structure, which is clearly indicates by high V$_{th}$ of MAPbBr$_3$ based devices after pre-biasing the deice at high V$_b$.

To more accurately determine the built-in voltage (V$_{bi}$) in our devices we conducted capacitance voltage characteristics of dual functional devices in LED regime. Fig.3a demonstrates the obtained C-V characteristics of dual functional devices based on MAPbBr$_2$I (Fig. S8). The black curve corresponds to unbiased device and the red to the device biased at 1.5V for 1 minute. The presence of a "kink" on the red curve corresponds to charge accumulation in the device [45]. The Mott–Schottky analysis [46] of obtained C-V characteristics for devices based on MAPbBr$_2$I, FAPbBr$_2$I, and MAPbBr$_3$ gave values of $V_{bi}$ around 1.7, 2.6 and 2.9 V (Fig. 3b), which are in agreement with those obtained from LED J-V characteristics [45] (Figs. 2 and 3).

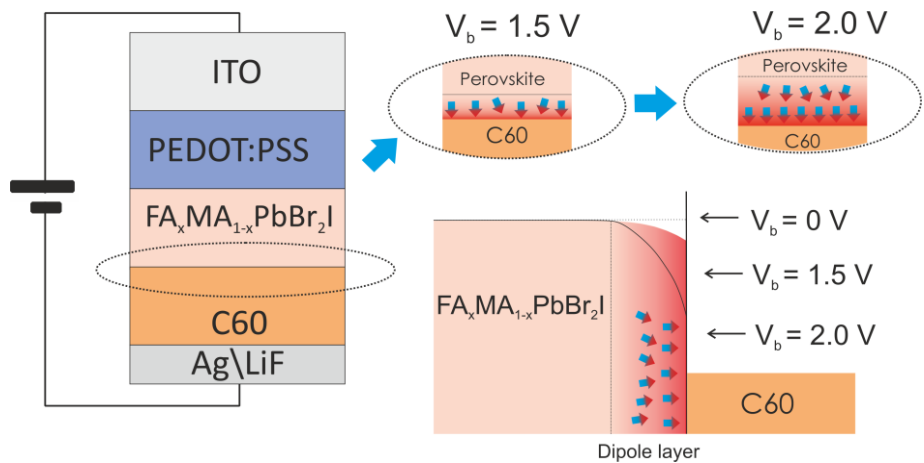

**Figure 4. Model. a** Cation migration during the device pre-biasing (anion ions omitted for simplicity). The pre-biasings at different $V_b$ lead to accumulation of cation ions at the perovskite\ETL layer. Higher $V_b$ leads to higher cation density at the interface. As a result device will have different $V_{th}$ values because different cation densities produce different device band bending.

According to latest investigations of halide diffusion in MAPbI$_3$ the cation ion is almost intact as the iodine ion moves across the perovskite [33, 34]. Apparently, mixed halide perovskites demonstrate both cation and anion ions migration under applied external voltage. And pre-biasing plays important role in achievement of low $V_{th}$ of the LED regime. During pre-biasing under applied voltage anion and cation ions move toward the perovskite/transport layer interfaces and form accumulation layers. The cation accumulation forming at the perovskite/C$_{60}$ layer leads to band bending due to its non-zero dipole momentum. In this case, the cation accumulation layer works as a PEI layer [31,39]. A higher value of $V_b$ results in a lower $V_{th}$ (Fig 3a) due to the formation of a stronger accumulation layer (Fig. 4). Since MA$^+$ and FA$^+$ have different dipole momentums, this requires different $V_b$ needed to bend band diagram and achieve low $V_{th}$ (Fig. 2 and 4). Moreover, dipole momentum of MA molecule is comparable with dipole momentums of ETL widely used in organic light emitting diodes like BCP and TPBI [45]. Therefore, ionic migration of both cation and anion ions of perovskite along with high dipole momentum of organic cation make mixed halide perovskite more perspective than monohalide ones for development of LESC.

## Conclusions

We have shown that the ionic migration in mixed halide perovskites demonstrates cation and anion movement, resulting in dipole layer formation near the perovskite/ETL interface, which lowers the threshold voltage for the LED operation mode. The presence of these accumulation layers leads to formation of more pronounced p-i-n structure inside the perovskite layer during device pre-biasing. Since molecules of perovskite organic cations MA and FA have non-zero dipole momentums of 2.29 and 0.21 Debye, respectively, band bending allows us to overcome high potential barrier between perovskite and transport layer in standard solar cell design based on PEDOT:PSS and C$_{60}$ as transport layers. In case of MAPbBr$_2$I perovskite, the $V_{th}$ value can be as low as 1.7V, whereas the FAPbBr$_2$I perovskite gives 2.6V, and MAPbBr$_3$ perovskite barely reaches 2.9V. Thus, the device band diagram can be switched from solar cell to LED operation, and therefore creating the dual functional device mixed halide perovskites have more perspective than monohalide ones due to both cation and anion migration.

# References

[1] Lu M., Zhang Y., Wang S., Guo J., Yu W. W., & Rogach A. L. Metal Halide Perovskite Light-Emitting Devices: Promising Technology for Next-Generation Displays. *Advanced Functional Materials*, **2019**, 1902008.

[2] Liu Z., Krückemeier L., Krogmeier B., Klingebiel B., Márquez J. A., Levcenko S., Oz S., Mathur S., Rau U., Unold T., Kirchartz T. Open-circuit voltages exceeding 1.26 v in planar methylammonium lead iodide perovskite solar cells. *ACS Energy Letters*, **2018**, 4, 110-117.

[3] Stranks S. D., Hoye R. L., Di D., Friend R. H., Deschler F. The physics of light emission in halide perovskite devices. *Advanced Materials*, **2018** 1803336.

[4] Jung M., Ji S. G., Kim G., Seok S. I. Perovskite precursor solution chemistry: from fundamentals to photovoltaic applications. *Chemical Society Reviews*, **2019**, 48, 2011-2038.

[5] Becker M. A., Vaxenburg R., Nedelcu G., Sercel P. C., Shabaev A., Mehl M. J., Michopoulos J. G. , Lambrakos S. G. , Bernstein N., Lyons J.L. , Stöferle T., Mahrt R.F. , Kovalenko M.V., Norris D.J. , Rainò G. and Efros A.L. Bright triplet excitons in caesium lead halide perovskites. *Nature*, **2018,** 553, 189.

[6] Yang W. S., Park B. W., Jung E. H., Jeon N. J., Kim Y. C., Lee D. U., S. S. Shin, J. Seo, E. K. Kim, J. H. Noh and Seok S. I. Iodide management in formamidinium-lead-halide–based perovskite layers for efficient solar cells. *Science*, **2017**, 356, 1376-1379.

[7] McMeekin D.P., Sadoughi G., Rehman W., Eperon G.E., Saliba M., Hörantner M.T., Haghighirad A., N. Saka1, L. Korte, B. Rech, M.B. Johnston, L.M. Herz, H.J. Snaith. A mixed-cation lead mixed-halide perovskite absorber for tandem solar cells. *Science*, **2016**, 351, 151-155.

[8] Seok S.I., Grätzel M. and Park N., Methodologies toward Highly Efficient Perovskite Solar Cells. *Small*, **2018**, 14, 1704177.

[9] Lin K., Xing J., Quan L.N., de Arquer F.P.G., Gong X., Lu J., Xie L., Zhao W., Zhang D., Yan C., Li W., Liu X., Lu Yan, Kirman J., Sargent E.H., Xiong Q. and Wei Z. Perovskite light-emitting diodes with external quantum efficiency exceeding 20 per cent. *Nature*, **2018**, 562, 245.

[10] Brennan M.C., Draguta S., Kamat P.V and Kuno M., Light-Induced Anion Phase Segregation in Mixed Halide Perovskites. *ACS Energy Lett.*, **2018**, 3, 204–213.

[11] Yuan Y. and Huang J. Ion Migration in Organometal Trihalide Perovskite and Its Impact on Photovoltaic Efficiency and Stability. *Acc. Chem. Res.*, 2016, 49, 286–293.

[12] Hoke E. T., Slotcavage D. J., Dohner E. R.,. Bowring A. R, Karunadasa H. I. and McGehee M. D. Reversible photo-induced trap formation in mixed-halide hybrid perovskites for photovoltaics. *Chem. Sci.*, **2015**, 6, 613–617.

[13] Braly I. L., Stoddard R. J., Rajagopal A., Uhl A. R., Katahara J. K., Je, A. K. Y. and Hillhouse H. W. Current-induced phase segregation in mixed halide hybrid perovskites and its impact on two-terminal tandem solar cell design. *ACS Energy Letters*, **2017**, 2, 1841-1847.

[14] Slotcavage D. J., Karunadasa H. I. and McGehee M. D. Light-Induced Phase Segregation in Halide-Perovskite Absorbers. *ACS Energy Lett.*, **2016** 1, 1199–1205.

[15] Knight A. J., Wright A. D., Patel J. B., McMeekin D. P., Snaith H. J., Johnston M. B. and Herz

L. M. Electronic Traps and Phase Segregation in Lead Mixed-Halide Perovskite. *ACS Energy Letters*, **2018**, 4, 75-84.

[16] Aristidou N., Eames C., Sanchez-Molina I., Bu X., Kosco J., Islam M. S. and Haque S. A. Fast oxygen diffusion and iodide defects mediate oxygen-induced degradation of perovskite solar cells. *Nature communications*, **2017**, 8, 15218.

[17] Besleaga C., Abramiuc L. E., Stancu V., Tomulescu A.G., Sima M., Trinca L., Plugaru N., Pintilie L. Nemnes G.A., Iliescu M., Svavarsson H.G., Manolescu A., Pintilie I. Iodine migration and degradation of perovskite solar cells enhanced by metallic electrodes. *The journal of physical chemistry letters*, **2016**, 7, 5168-5175

[18] Eames C., Frost J. M., Barnes P. R. F., O'Regan B. C., Walsh A. and Islam M. S. Ionic transport in hybrid lead iodide perovskite solar cells. *Nat. Commun.*, **2015**, 6, 2–9.

[19] Smith E. C., Ellis C. L., Javaid H., Renna L. A., Liu Y., Russell T. P., Bag M. and Venkataraman D. Interplay between Ion Transport, Applied Bias, and Degradation under Illumination in Hybrid Perovskite pin Devices. *The Journal of Physical Chemistry C*, **2018**, 122, 13986-13994.

[20] Habisreutinger S. N., Noel N. K. and Snaith H. J. Hysteresis Index: A Figure without Merit for Quantifying Hysteresis in Perovskite Solar Cells. *ACS Energy Lett.*, **2018**, 3, 2472–2476.

[21] Ferdani D., Pering S., Ghosh D., Kubiak P., Walker A., Lewis S. E., Johnson A.L., Bakedr P.J., Islam M.S. and Cameron P. J. Partial Cation Substitution Reduces Iodide Ion Transport in Lead Iodide Perovskite Solar Cells. *Energy & Environmental Science*, **2019**, 12, 2264-2272.

[22] Yoo J.J., Wieghold S., Sponseller M., Chua M., Bertram S. N., Hartono N.T.P., Tresback J.S., Hansen E.C., Correa-Baena J.-P., Bulović V., Buonassisi T., Shin S.S. and Bawendi M.G. An Interface Stabilized Perovskite Solar Cell with High Stabilized Efficiency and Low Voltage Loss. *Energy & Environmental Science*, **2019**, 12, 2192-2199

[23] Han T. H., Tan S., Xue J., Meng L.,. Lee J. W, and Yang Y. Interface and Defect Engineering for Metal Halide Perovskite Optoelectronic Devices. *Adv. Mater.*, **2019**, 1803515, 1–35.

[24] Hu M., Bi C., Yuan Y., Bai Y., and Huang J. Stabilized wide bandgap MAPbBrxI3-x perovskite by enhanced grain size and improved crystallinity. *Adv. Sci.*, **2015**, 3, 6–11.

[25] Walsh A. and Stranks S. D. Taking Control of Ion Transport in Halide Perovskite Solar Cells. *ACS Energy Lett.*, **2018**, 3, 1983–1990.

[26] Deng Y., Xiao Z. and Huang J. Light-Induced Self-Poling Effect on Organometal Trihalide Perovskite Solar Cells for Increased Device Efficiency and Stability. *Adv. Energy Mater.*, **2015**, 5, 1500721.

[27] Lee J.-W., Kim S.-G., Yang J.-M., Yang Y., and Park N.-G.. Verification and mitigation of ion migration in perovskite solar cells. *APL Mater.*, **2019**, 7, 4, 041111.

[28] Puscher B. M. D., Aygüler M. Docampo F., P., and Costa R. D. Unveiling the Dynamic Processes in Hybrid Lead Bromide Perovskite Nanoparticle Thin Film Devices. *Adv. Energy Mater.*, **2017**, 7, 1–10.

[29] Zhang T., Cheung S. H., Meng X., Zhu L., Bai Y., Ho C. H. Y., Xiao S., Xue Q. So S.K. and Yang, S. Pinning down the anomalous light soaking effect toward high-performance and fast-response perovskite solar cells: the ion-migration-induced charge accumulation. *The journal of physical chemistry letters*, **2017** 8, 5069-5076.

[30] Gets D., Saranin D., Ishteev A., Haroldson R., Danilovskiy E., Makarov S., and Zakhidov A. Light-emitting perovskite solar cell with segregation enhanced self doping. *Applied Surface Science*, **2019** 476, 486-492.

[31] Kim H. B., Yoon Y. J., Jeong J., Heo J., Jang H., Seo J. H., Walker B., Kim J. Y. Peroptronic devices: perovskite-based light-emitting solar cells. Energy & Environmental Science, **2017**, 10, 1950-1957.

[32] Gil-Escrig L., Longo G., Pertegás A., Roldán-Carmona C., Soriano A., Sessolo M. and Bolink H. J. Efficient photovoltaic and electroluminescent perovskite devices. *Chemical Communications*, **2015**, 51, 569-571.

[33] Senocrate A., Moudrakovski I., Acartürk T., Merkle R., Kim G. Y., Starke U., Grätzel M. and Maier J. Slow CH3NH3+ diffusion in CH3NH3PbI3 under light measured by solid-state NMR and tracer diffusion. *The Journal of Physical Chemistry C*, **2018**, 122, 21803-21806.

[34] Futscher M. H., Lee J. M., McGovern L., Muscarella L. A., Wang T., Haider M. I., Fakharuddin A., Schmidt-Mende L., Ehrler B. Quantification of ion migration in CH3NH3PbI3 perovskite solar cells by transient capacitance measurements. *Materials Horizons*, **2019**, 6, 1497-1503.

[35] Chiba T., Kumagai D., Udagawa K., Watanabe Y. and Kido J. Dual mode OPV-OLED device with photovoltaic and light-emitting functionalities. *Scientific reports,* **2018**, 8, 11472.

[36] Aygüler M. F., Puscher B. M., Tong Y., Bein T., Urban A. S., Costa R. D., and Docampo P. Light-emitting electrochemical cells based on inorganic metal halide perovskite nanocrystals. Journal of Physics D: Applied Physics, **2018**, 51, 334001.

[37] Liu Y., Hangoma P. M., Tamilavan V., Shin I., Hwang I. W., Jung Y. K., Lee B.R., Jeong J.H., Park S.H., Kim K. H. Dual-functional light-emitting perovskite solar cells enabled by soft-covered annealing process. *Nano Energy*, **2019**, 61, 251-258.

[38] Frost J. M., Butler K. T., Brivio F., Hendon C. H., van Schilfgaarde M. and Walsh A. Atomistic Origins of High-Performance in Hybrid Halide Perovskite Solar Cells. *Nano Lett.*, **2014**, 14, 2584–2590.

[39] Zhou Y., Fuentes-Hernandez C., Shim J., Meyer J., Giordano A. J., Li H., Winget P., Papadopoulos T., Cheun H., Kim J., Fenoll M., Dindar A., Haske W., Najafabadi E., Khan T.M., Sojoudi H., Barlow S., Graham S., Brédas J.-L., Marder S.R., Kahn A., Kippelen B. A universal method to produce low–work function electrodes for organic electronics. *Science*, **2012**, 336, 327-332.

[40] Bernard G. M., Wasylishen R. E., Ratcliffe C. I., Terskikh V., Wu Q., Buriak J. M. and Hauger T. Methylammonium Cation Dynamics in Methylammonium Lead Halide Perovskites: A Solid-State NMR Perspective. *The Journal of Physical Chemistry A*, **2018**, 122, 1560-1573.

[41] Yin W.-J., Shi T., and Yan Y. Unusual defect physics in CH3NH3PbI3 perovskite solar cell absorber. *Appl. Phys. Lett.*, 2014, 104, 063903.

[42] Liu N. and Yam C. First-principles study of intrinsic defects in formamidinium lead triiodide perovskite solar cell absorbers. *Phys. Chem. Chem. Phys.*, **2018**, 20, 6800–6804.

[43] Kang J. and Wang L. High Defect Tolerance in Lead Halide Perovskite CsPbBr3. *J. Phys. Chem. Lett.*, **2017**, 8, 489–493.

[44] Jiang Q., Chen M., Li J., Wang M., Zeng X., Besara T., Lu J., Xin Y., Shan X., Pan B., Wang C.,

Lin S., Siergrist T., Xiao Q., Yu Z. Electrochemical doping of halide perovskites with ion intercalation. *ACS nano*, **2017**, 11, 1073-1079.

[45] Noguchi Y., Miyazaki Y., Tanaka Y., Sato N., Nakayama Y., Schmidt T. D., Brütting W., and Ishii H. Charge accumulation at organic semiconductor interfaces due to a permanent dipole moment and its orientational order in bilayer devices. *Journal of Applied Physics*, **2012**, 111, 114508.

[46] Almora O., Aranda C., Mas-Marzá E. and Garcia-Belmonte G. On Mott-Schottky analysis interpretation of capacitance measurements in organometal perovskite solar cells. *Applied Physics Letters*, **2016**, 109, 173903.